\documentclass[preprint]{vgtc}                          

\graphicspath{{figures/}{pictures/}{images/}{./}} 

\usepackage{times}                     

\usepackage{tabu}                      
\usepackage{booktabs}                  
\usepackage{lipsum}                    
\usepackage{mwe}                       

\usepackage{mathptmx}                  

\usepackage{xargs} 
\usepackage{soul}  
\usepackage{xspace}
\usepackage{listings}

\usepackage[vflt]{floatflt}

\newcommand{\ie}{{i.e.,}\xspace}
\newcommand{\eg}{{e.g.,}\xspace}
\newcommand{\cf}{{c.f.}\xspace}
\newcommand{\ea}{{et~al.}\xspace}

\newcommand{\bpstart}[1]{\vspace{1mm} \noindent{\textbf{#1.}}}

\onlineid{0}

\vgtccategory{Research}

\vgtcinsertpkg

\title{Bringing Data into the Conversation: Adapting Content from Business Intelligence Dashboards for Threaded Collaboration Platforms}

\author{
Hyeok Kim\thanks{e-mail: hyeok@northwestern.edu}\\ %
        \scriptsize Northwestern University %
\and Arjun Srinivasan\thanks{e-mail: arjunsrinivasan@tableau.com}\\ %
     \scriptsize Tableau Research %
\and Matthew Brehmer\thanks{e-mail: mbrehmer@uwaterloo.ca}\\ %
     \scriptsize University of Waterloo
}

\abstract{
To enable data-driven decision-making across organizations, data professionals need to share insights with their colleagues in context-appropriate communication channels.
Many of their colleagues rely on data but are not themselves analysts; furthermore, their colleagues are reluctant or unable to use dedicated analytical applications or dashboards, and they expect communication to take place within threaded collaboration platforms such as Slack or Microsoft Teams.
In this paper, we introduce a set of six strategies for adapting content from business intelligence (BI) dashboards into appropriate formats for sharing on collaboration platforms, formats that we refer to as \textit{dashboard snapshots}. 
Informed by prior studies of enterprise communication around data, these strategies go beyond redesigning or restyling by considering varying levels of data literacy across an organization, introducing affordances for self-service question-answering, and anticipating the post-sharing lifecycle of data artifacts.
These strategies involve the use of templates that are matched to common communicative intents, serving to reduce the workload of data professionals.
We contribute a formal representation of these strategies and demonstrate their applicability in a comprehensive enterprise communication scenario featuring multiple stakeholders that unfolds over the span of months. 
} 


\begin{document}

\definecolor{ttBlue}{cmyk}{1, 0.5, 0, 0.5}
\newcommand{\ttt}[1]{\textcolor{ttBlue}{\texttt{#1}}}

\firstsection{Introduction}

\maketitle

\noindent
Distributed and asynchronous workplace communication is increasingly prevalent in organizations, from enterprise corporations to government agencies and academic institutions. 
One indicator of this shift is the increased adoption of collaboration platforms~\cite{wang2022groupchat} such as Slack~\cite{slackStat} or Microsoft Teams~\cite{teamsStat}.
While these platforms offer convenient affordances for threaded conversations and file sharing along with extensions for productivity applications, collaboration around data on these platforms can be challenging~\cite{zhang2020data}.

Early work on collaborative visualization~\cite{heer2007design,isenberg2011} envisioned ways to bring asynchronous collaboration to visual analytic tools, such as within business intelligence (BI) applications like Tableau~\cite{tableau} or Power BI~\cite{powerbi}. 
However, recent research on enterprise communication around data~\cite{brehmer2022,tory2021datavoice} reveals barriers and aversions to context switching, and that it is tedious to generate and share static data artifacts (\eg~presentations and screenshots) across organization's communication channels.
Unfortunately, these artifacts may not serve audience with varying levels of data literacy, raise data governance concerns, and fail to satisfy needs for simple follow-up questions.
Instead of adding collaborative affordances to dashboards and BI applications, our work considers ways to adapt content from dashboards for collaboration platforms while respecting considerations for data literacy, governance, and a desire for self-service question-answering. 

The primary contribution of this work is a set of six strategies, each supported by formal representations, for systematically adapting dashboard content for sharing on an organization's collaboration platform. 
Inspired by recent formal, concise specification of dashboard elements~\cite{epperson2023:quickDash,heer2024:masaic,mcnutt:2021ivy,wongsuphasawat:2020encodable}, our strategies extend this form of expression to encompass the selection and adaptation of dashboard content as a \textit{dashboard snapshot}, referring to the state of a selection made from a dashboard at a certain point in time. 
We showcase these strategies in a comprehensive demonstration scenario of asynchronous communication involving multiple stakeholders across an organization over months.
For this scenario, we built a graphical user interface to serve as an intermediary between a BI dashboard and a collaboration platform, allowing data professionals to specify snapshots in accordance with our proposed strategies.

\section{Background and Prior Work}
\label{sec:rw}

\bpstart{Collaborative visualization in enterprise organizations} 
Isenberg~\ea~\cite{isenberg2011} characterize collaborative visualization in terms of time (synchronous vs. asynchronous) and place (co-located vs. distributed).
We focus on asynchronous distributed collaboration within organizations.
To support collaborative data analysis, prior work~\cite{heer2007design,heer2009,CommentSpace} suggests methods to extend visual analysis tools, such as by allowing for making threaded conversation and drawing attention to insights via annotation.
However, within enterprise organizations, these techniques are difficult to implement as many individuals lack access and orientation to BI applications~\cite{tory2021datavoice} such as Tableau~\cite{tableau} or Power BI~\cite{powerbi}.

People communicate \textit{with and around} data~\cite{tory2021datavoice} in organizations of varying sizes to support data-driven decision making~\cite{crisan2021,zhang2020data}.
Prior work~\cite{kandogan2014,kandel2012,zhang2020data} identifies two primary roles in enterprise data communication: \textit{analysts} who work with the data and share artifacts with \textit{consumers}.
Consumers expect communication to take place on collaboration platforms.
As a consequence, analysts are tasked with capturing, \textit{`slicing and dicing,'} or painstakingly reproducing content from dashboards~\cite{tory2021datavoice} and sharing them as screenshots or presentation slides~\cite{brehmer2022}, suggesting tedious workflows that require switching between several tools.
As these artifacts are static, this process must be repeated when dashboards are modified or when new data becomes available.
Additionally, these static artifacts will not observe BI applications' data governance policies for ensuring appropriate role-based access to data.
If conversations around data are to take place on collaboration platforms, our proposed strategies respect both a desire for self-service question-answering among consumers as well as the needs of analysts to streamline and govern communication workflows. 

\bpstart{Visualization retargeting}
Visualization retargeting is a general term that encompasses both responsive visualization design across devices~\cite{hoffswell2020,kim:responsive2021} as well as restyling chart designs~\cite{Harper2014}.
However, retargeting often entails much more than resizing or restyling content; it requires an understanding of the communicative intent of the content and which elements are essential to retain ~\cite{bach:2022dashboardDesignPattern,digiacomo2015:network,kim:responsive2021,kim:insight2021,wu2013:visizer}.
Here, we use the term to describe the retargeting of visualization from a dashboard context to a collaboration platform context.
This change of context also entails a change of audience: from analysts engaging with the content and organization of a dashboard to consumers with varying levels of data literacy.
While BI applications offer features for bookmarking parts of a dashboard~\cite{powerbibookmark,tableauMetrics} or making a data story~\cite{powerbistory,tableauStories}, they do not necessarily support retargeting dashboard contents for message-based asynchronous conversations.

\bpstart{Expressing visualization design}
Declarative specifications of visualization design are foundational in building useful visual analysis interfaces.
For example, Vega-Lite~\cite{satyanarayan:vega-lite2017} and ggplot2~\cite{hadley2010:ggplot2} implement the \textit{Grammar of Graphics}~\cite{wilkinson2012:grammar}, expressing a visualization design as a set of encoding channels like \textit{x}, \textit{y}, and \textit{color}.
Ivy~\cite{mcnutt:2021ivy} provides a method to templatize visualization design specs using such grammars to enhance their reusability. 
Similarly, Encodable~\cite{wongsuphasawat:2020encodable} offers a way to configure visualization templates for reactive web environments (\eg~React), and
Quick Dashboard~\cite{epperson2023:quickDash} provides lightweight expressions for configuring dashboard elements using analytic terms like `Metrics' and `Dimensions.'
However, while these template-based approaches are suitable for specifying dashboards, they do not include considerations for constructing visual analysis experiences within the context of collaboration platforms.
Our work therefore introduces strategies with corresponding declarative representations for expressing \textit{dashboard snapshots} in this collaborative context, strategies that consider the ease of retargeting and the life cycle of artifacts on such platforms.
\section{Design Considerations}
\label{sec:design}

Motivated by prior work on enterprise communication around data~\cite{crisan2021,donoho2017,isenberg2011,kandel2012,kandogan2014,tory2021datavoice,zhang2020data}, our proposed strategies observe the following high-level considerations for addressing the communication gap between BI applications and collaboration platforms.

\bpstart{(DC1) Be specific and transparent}
Analysts often focus on a certain part of the data shown in a dashboard by selecting a few fields or a subset of records~\cite{tory2021datavoice}.
Information about such a selection as well as data incompleteness (\eg~missing data) must be transparent to not mislead consumers~\cite{kandel2012}.

\bpstart{(DC2) Be context-appropriate}
We must adapt selected content from a dashboard in ways that make it appropriate for and complementary to conversations around data.
Retargeted dashboard snapshots should be comprehensible for those with limited visualization literacy, which may entail the generation of text-based summaries~\cite{kandel2012} or the addition of annotations~\cite{heer2009}.
Furthermore, a level of interactivity may need to be retained, so as to support consumers as they form predictable questions of their own (\eg~\textit{``what is the trend for each product category?''})~\cite{kandel2012}.

\bpstart{(DC3) Be time-sensitive}
Relying on stale data (such as screenshots of dashboards) in an organizational setting can lead to erroneous decisions, so analysts often need to produce updated artifacts.
While some updates can be automated, others require validation, such as verifying the validity of new data or identifying contextual factors affecting snapshots' quality~\cite{crisan2021,kandel2012}.

\begin{figure}[b!]
    \vspace{-1em}
    \centering
    \includegraphics[width=\columnwidth]{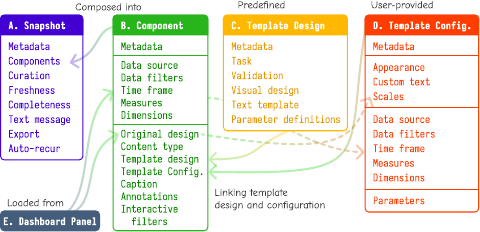}
    \caption{The relationships between snapshots (A), components (B), templates (C), template specifications (D), and dashboard selections (E). Components load data, measures, breakdowns, filters, and worksheet from the original dashboard selection, which are propagated to a template specification for a template-based component.}
    \label{fig:formal}
\end{figure}

\begin{figure}[t!]
    \centering
    \includegraphics[width=\columnwidth]{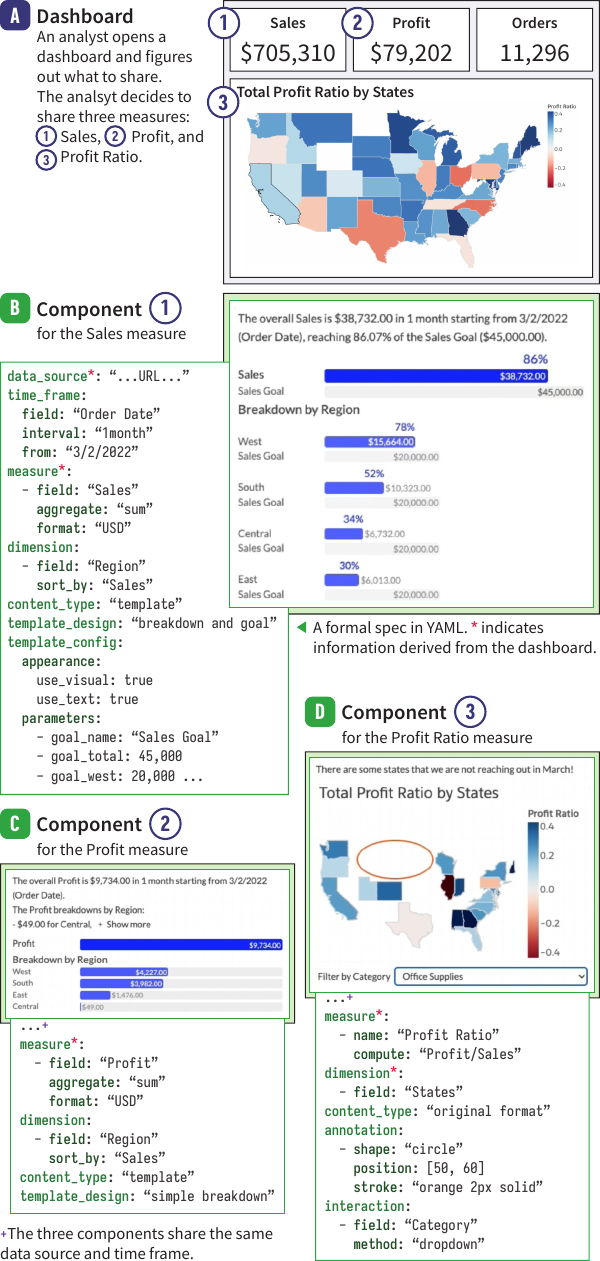}
    \caption{Example components made from dashboard selections and their formal representations in YAML. 
    }
    \label{fig:scenario}
    \vspace{-2em}
\end{figure}

\section{Six Strategies for Adapting Dashboard Content}
\label{sec:strategies}

\begin{figure}[t!]
    \centering
    \includegraphics[width=\columnwidth]{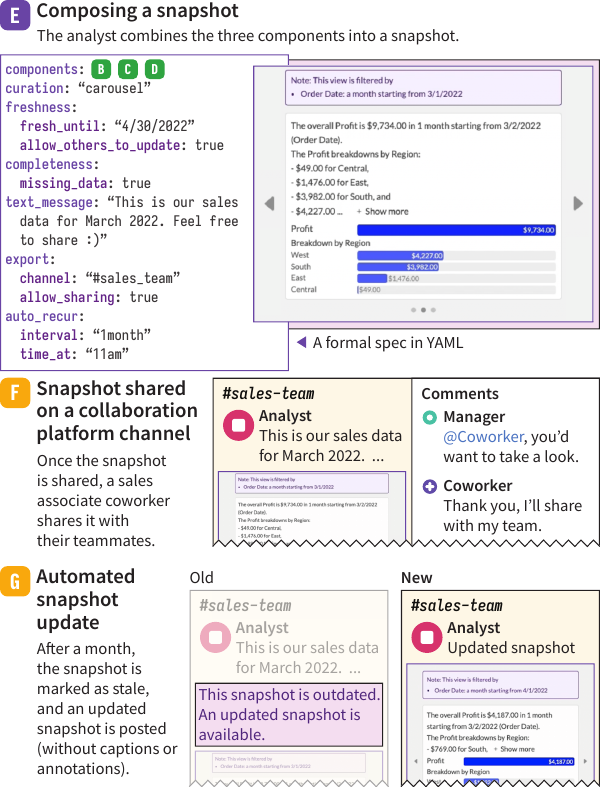}
    \caption{An example snapshot composed of the components in \autoref{fig:scenario} and its formal representation in YAML.}
    \label{fig:scenario2}
    \vspace{-2em}
\end{figure}

We define a \textbf{dashboard snapshot} as an artifact that communicates a selection of one or more elements from a dashboard at a given point in time, making an analogy to snapshots used in cloud data platforms (\eg~\cite{nces,oracle}).
Motivated by the above considerations, we describe strategies for systematically adapting dashboard selections to snapshots, accompanied by formal representations for elements listed in \autoref{fig:formal}.
\autoref{fig:scenario} and \autoref{fig:scenario2} illustrate example snapshot elements along with their formal representations in YAML.

\bpstart{Strategy 1: Anticipate selections of varying scope}
As the scope of a dashboard snapshot must be flexible (\textbf{DC1}), a snapshot can have multiple \textit{component}s, each having a different selection.
As shown in \autoref{fig:scenario}A, a dashboard is typically arranged into zones or panels, with any individual panel containing a chart or a metric; panels generally show measures (quantitative variables) and dimensions (nominal or temporal variables).
An analyst can create a \ttt{component} from a panel (\autoref{fig:scenario}B--D).
The \ttt{component} imports the \ttt{data source}, \ttt{data filters}, \ttt{measures}, \ttt{dimensions} and \ttt{original design} from the selected element (\autoref{fig:formal}B).
However, it is not necessary to include all of these in a single component.
If the analyst wants to include information from other panels (\textbf{DC1}), they can opt to create a multi-component snapshot.
As a measure in a dashboard can be traced back to a column in the data, a newly computed variable from multiple measures (\autoref{fig:scenario}D), or an aggregated variable (\autoref{fig:scenario}B \& C), a \ttt{measure} in a \ttt{component} must reflect this potential variability.
When importing \ttt{dimensions} and \ttt{data filters}, a \ttt{component} must identify a \ttt{time frame} that will to be used to infer the temporal validity of a snapshot (\textbf{DC3}). 
For example, the component in \autoref{fig:scenario}B for sales data has a time frame of `1 month' from March 2, 2022 by the `Order Date' field.
If the original panel lacks an explicit temporal variable, an analyst can impose their own time frame. 
Both the \ttt{data filters} and the \ttt{time frame} should appear within the component for transparency (\textbf{DC1}, see \autoref{fig:scenario2}E).

\bpstart{Strategy 2: Provide affordances for adding situational context}
The analyst can add a \ttt{caption} (plain text) and \ttt{annotations} (visual emphases ovarlaid on a chart).
To enable simple and predictable question-answering within a conversation thread, an analyst can set \ttt{interactive filters} with drop-down menus (\autoref{fig:scenario}D) or sliders.
Alternatively, the analyst can apply combinations of filters as a `macro,' so that consumers can reproduce transformations and contrast the results against unfiltered data themselves.
When a consumer applies a component's interactive filter, this transformation should be private view to that consumer, so as not to mislead the other consumers (\textbf{DC2}).

\bpstart{Strategy 3: Offer templates alongside original designs}
The analyst can opt to keep the \ttt{original design} (\autoref{fig:scenario}D) or use a template for predictable and recurring communicative intents (\autoref{fig:scenario}B \& E) (\textbf{DC2}). 
For the former, the extent of adaptation should not exceed minimal responsive transformations (\eg~changes in aspect ratio and font size, removing axis ticks; \cf~\cite{hoffswell2020,kim:responsive2021}).
For the latter, the analyst can choose a \ttt{template design} and specify its \ttt{template config}uration. 
Each template must have a clear communicative intent (\textbf{DC2}).
As illustrated in \autoref{fig:scenario}B, for example, a \ttt{template design} for \textit{`breakdown and goal'} communicates whether a measure has met a threshold value (\eg~a sales target).
Like Tableau's Show Me~\cite{showme}, a snapshot authoring interface should validate the applicability of templates for a component according to the contents of the selected dashboard panel.

The \ttt{visual design} of a template can assume various formats, and may or may not include a chart, and a component's \ttt{appearance} property lets the analyst opt to use the \ttt{visual design}, the caption \ttt{text template}, or both. 
Caption \ttt{text template} (\ie parameterized sentences) can assist consumers with limited visualization literacy (\textbf{DC2}).
A \ttt{custom text} (as a plain text or a parameterized expression) allows for overriding the \ttt{text template}.
Continuing our second strategy, a template can allow for the inclusion of values that were not part of the selected dashboard panel, and thus a template design can have \ttt{parameter definitions}.
For example, for the template design shown in \autoref{fig:scenario}B, goal parameters are included for each category of a dimension, and these goal values are rendered as gray bars in the \ttt{visual design} and ratios in the caption \ttt{text template}.

\begin{figure}[!t]
  \centering
  \includegraphics[width=\linewidth]{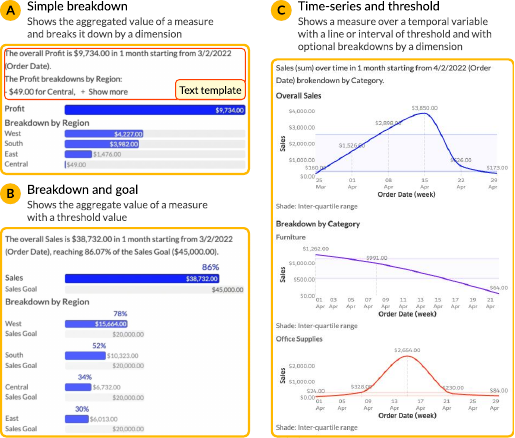}
  \caption{Example templates for simple breakdown (A), a breakdown with a goal (B), and a time-series with a threshold (C).}
  \label{fig:template}
  \vspace{-2em}
\end{figure}

Templates should reflect common communicative intents, such as those classified in Bach~\ea's dashboard design patterns~\cite{bach:2022dashboardDesignPattern}.
For example, \autoref{fig:template}A shows a template for the simple breakdown of a measure by a dimension, and \autoref{fig:template}B adds progress information to that by defining `goal' parameters for total and each dimension value. 
\autoref{fig:template}C presents a template for the time-serial trend of a measure, which can optionally be disaggregated by a dimension or include upper and lower threshold parameter values.

A \ttt{template config}uration object (\autoref{fig:formal}D) mediates the selected \ttt{template design} for a \ttt{component}, inspired by Encodable~\cite{wongsuphasawat:2020encodable}.
Most of its attributes (\eg~measures, time frame) are directly transferred from the component.
The \ttt{scales} are transferred from the \ttt{original design} to preserve color legends or scale transformations.
Lastly, adding template \ttt{parameter} values to this configuration is the responsibility of the analyst.

\bpstart{Strategy 4: Support curation of multi-component snapshots}
For snapshots with more than one \ttt{component}, the analyst can choose a \ttt{curation} method for \ttt{components} that are suitable for sharing on threaded collaboration platforms (\textbf{DC2}).
As illustrated in \autoref{fig:scenario2}E, these curation methods can include vertical stacking, a component carousel, an automated slideshow, or a mini-dashboard.

\bpstart{Strategy 5: Convey data caveats transparently}
The analyst can set \ttt{freshness} and \ttt{completeness} to communicate the temporal validity and collection reliability of the data being represented (\textbf{DC1} \& \textbf{DC3}).
A \ttt{freshness} property is akin to a \textit{best-before date} that annotates the snapshot.
This date can be inferred from the \ttt{time frame}s of the snapshot's \ttt{components}.
A default inference method is to find the \ttt{time frame} with the latest end date and add the \ttt{time frame}'s duration to the end date, so that the snapshot can wait for the data for the next period to be fully collected.

The analyst can also indicate the existence of missing data points with a \ttt{completeness} indicator.
To provide additional context to either a component's \ttt{freshness} or \ttt{completeness}, an analyst can add a \ttt{text message} to the snapshot.

\bpstart{Strategy 6: Plan to update snapshots}
When a snapshot becomes stale (\ie~after its \ttt{freshness} date), it should be clearly indicated so as to prevent people from relying on old data (\textbf{DC3}). 
Then, the old snapshot may need to be updated with new data (\textbf{DC3}).
A snapshot can be updated in three ways: (1)~manually updated by the snapshot author, (2)~manually updated by a viewer, and (3)~automatically updated according to a schedule set by the author (\ttt{auto-recur}).
To enable the latter, the analyst needs to provide the period of recurrence (\eg~1 month, 2 weeks), until when this recurrence should happen, and what time the updated snapshot is published at, similar to setting recurring events on a calendar. 

When a snapshot is updated, either via manual retrieval or an automatic recurrence, a system implementing this strategy must reference the \ttt{time frame} property of its components and adjust filters accordingly before posting the updated snapshot on a collaboration platform.
Additionally, updates to snapshots should exclude annotations from past snapshots so as not to mislead consumers.
If an analyst opts to manually update a snapshot, they can freely add annotations or captions before sharing the update.

\section{Demonstration Scenario}
\label{sec:demo}

\begin{figure}[t!]
    \centering
    \includegraphics[width=\columnwidth]{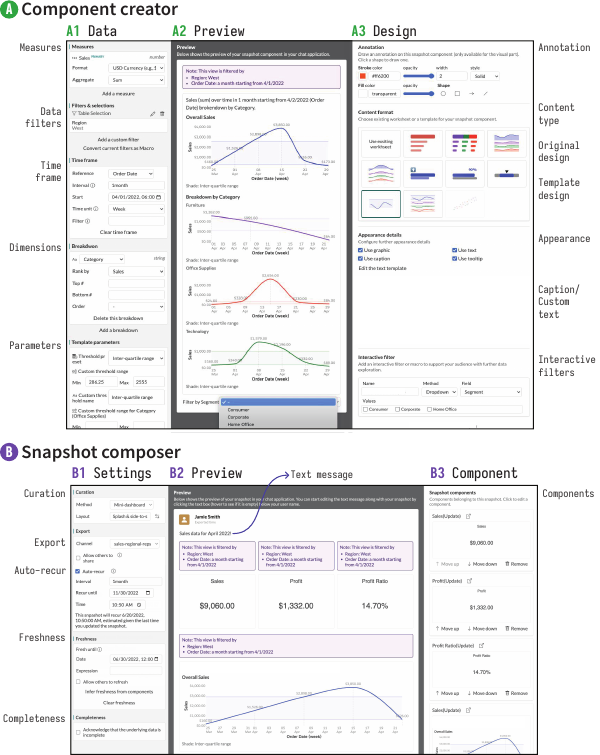}
    \caption{The Philo interface for specifying dashboard snapshots: (A) Component creator and (B) Snapshot composer. Formal properties are annotated next to the corresponding interface elements.}
    \label{fig:prototype}
    \vspace{-2em}
\end{figure}

To demonstrate the confluence of our strategies, we implemented an interactive BI dashboard and a threaded collaboration platform with basic functionalities similar to those of Tableau and Slack, respectively, along with \textbf{Philo}, an intermediary graphical user interface for specifying dashboard snapshots.
The Philo interface (\autoref{fig:prototype}) includes both a snapshot \textit{component creator} (a) as well as a \textit{snapshot composer} for producing multi-component snapshots (b).
We used Svelte~\cite{svelte} for the user interfaces, DynamoDB~\cite{dynamodb} for the data server, and Express~\cite{express} for the back-end server. 

We used the dashboard, collaboration platform, and Philo interfaces to produce a comprehensive demonstration scenario video in which a group of colleagues employed by an enterprise organization interact with dashboard snapshots over the course of a couple of months (see supplemental material). 
Specifically, the scenario describes how an analyst creates a snapshot in \autoref{fig:scenario} and \ref{fig:scenario2} and shares them with consumers, reifying \textbf{Strategies 1--5}. 
For the time-related behaviors, the scenario shows manual and automated updates (\textbf{Strategy 6}).
Finally, the scenario also encompasses how the analyst monitors the dissemination of snapshots across the channels of a collaboration platform over time. 
\section{Discussion, Limitations, and Future Work}
\label{sec:discussion}

We proposed six strategies for adapting BI dashboard content as snapshots on collaboration platforms, considering flexible content selection, conversations around snapshots, and time-sensitivity.
Through template-based retargeting and formal representation, these strategies have the potential to improve upon current  practices relating to conversations around data, obviating the need to create new dashboards or share screenshots of simple charts. 

While a human-centered usage evaluation could demonstrate the efficacy of our strategies, the collaborative and longitudinal nature of asynchronous enterprise conversations around data is methodologically and logistically challenging to capture in a single evaluation. 
This challenge is in part attributed to the need to study long-term collaborations between an analyst and multiple consumers, where the consumers exhibit a range of data literacy despite having situational context around the data. 
Individually, strategies 2 and 4 are applicable for focused studies with one-shot tasks; however, a holistic evaluation of all six strategies requires the deployment of an interface, such as Philo, that interoperates with production BI applications and collaboration platforms. 
To realize this, APIs for the former (\eg~the Tableau Extension API~\cite{tableauext}) would need to be extended to accommodate Strategies 1 and 4, while APIs for the latter (\eg~Slack's Block Kit~\cite{SlackBlockKit}, the Microsoft Graph API~\cite{TeamsGraphToolkit}) would need to be extended to accommodate Strategies 5 and 6. 
Once deployed, interesting future work includes identifying a more comprehensive set of component templates as well as organizational communication patterns arising from the dissemination of dashboard snapshots.

\section*{Supplemental Materials}
\label{sec:supplemental_materials}
We provide complete formal specifications for snapshot elements, a complete specification of proposed templates, as well as a video of our demonstration scenario.
It can also be accessed via \url{https://dashboard-snapshot.github.io}.
\section*{Acknowledgements}
\label{sec:ack}
The authors conducted this work while affiliated with Tableau.
\bibliographystyle{abbrv-doi}

\bibliography{template}
\end{document}